\documentclass[iop]{emulateapj}

\newcommand{\gps}{\ensuremath{g_{\rm P1}}}
\newcommand{\rps}{\ensuremath{r_{\rm P1}}}
\newcommand{\ips}{\ensuremath{i_{\rm P1}}}
\newcommand{\zps}{\ensuremath{z_{\rm P1}}}
\newcommand{\yps}{\ensuremath{y_{\rm P1}}}

\newcommand{\grizy}{\gps\rps\ips\zps\yps}

\def\ra#1#2#3{#1$^{\rm h}$#2$^{\rm m}$#3$^{\rm s}$}
\def\dec#1#2#3{$#1^\circ#2'#3''$}
\def\kms{km~s$^{-1}$}

\usepackage{epsfig,graphicx}
\usepackage{amsmath}
\usepackage{natbib}
\usepackage{comment}

\begin{document}

\title{PS1-10bzj: A fast, hydrogen-poor superluminous supernova in a metal poor host galaxy}

\submitted{ApJ in press}
\def\cfa{1}
\def\ifa{2}
\def\ucsc{3}
\def\qub{4}
\def\stsci{5}
\def\gemini{6}
\def\jhu{7}
\def\hup{8}
\def\pri{9}

\author{R. Lunnan\altaffilmark{\cfa},
 R.~Chornock\altaffilmark{\cfa},
 E.~Berger\altaffilmark{\cfa},
 D.~Milisavljevic\altaffilmark{\cfa},
 M.~Drout\altaffilmark{\cfa},
 N.~E.~Sanders\altaffilmark{\cfa},
 P.~M.~Challis\altaffilmark{\cfa},
 I.~Czekala\altaffilmark{\cfa},
 R.~J.~Foley\altaffilmark{\cfa},
 W.~Fong\altaffilmark{\cfa},
 M.~E.~Huber\altaffilmark{\ifa},
 R.~P.~Kirshner\altaffilmark{\cfa},
 C.~Leibler\altaffilmark{\cfa, \ucsc},
 G.~H.~Marion\altaffilmark{\cfa},
 M.~McCrum\altaffilmark{\qub},
 G.~Narayan\altaffilmark{\cfa},
 A.~Rest\altaffilmark{\stsci},
 K.~C.~Roth\altaffilmark{\gemini},
 D.~Scolnic\altaffilmark{\jhu},
 S.~J.~Smartt\altaffilmark{\qub},
 K.~Smith\altaffilmark{\qub},
 A.~M.~Soderberg\altaffilmark{\cfa},
 C.~W.~ Stubbs\altaffilmark{\hup},
 J.~L.~Tonry\altaffilmark{\ifa},
 W.~S.~Burgett\altaffilmark{\ifa}, 
 K.~C.~Chambers\altaffilmark{\ifa},
 R.-P.~Kudritzki\altaffilmark{\ifa},   
 E.~A.~Magnier\altaffilmark{\ifa}, and
 P.~A.~Price\altaffilmark{\pri} }

\altaffiltext{1}{Harvard-Smithsonian Center for Astrophysics, 60 Garden St., Cambridge, MA 02138, USA}
\altaffiltext{2}{Institute for Astronomy, University of Hawaii, 2680 Woodlawn Drive, Honolulu, HI 96822, USA}
\altaffiltext{3}{Department of Astronomy and Astrophysics, UCSC, 1156 High Street, Santa Cruz, CA 95064, USA}
\altaffiltext{4}{Astrophysics Research Centre, School of Mathematics and Physics, Queen's University Belfast, Belfast BT7 1NN, UK}
\altaffiltext{5}{Space Telescope Science Institute, 3700 San Martin Dr., Baltimore, MD 21218, USA} 
\altaffiltext{6}{Gemini Observatory, 670 N. Aohoku Place, Hilo, HI 96720, USA}
\altaffiltext{7}{Department of Physics and Astronomy, Johns Hopkins University, 3400 North Charles Street, Baltimore, MD 21218, USA}
\altaffiltext{8}{Department of Physics, Harvard University, Cambridge, MA 02138, USA} 
\altaffiltext{9}{Department of Astrophysical Sciences, Princeton University, Princeton, NJ 08544, USA} 

\email{rlunnan@cfa.harvard.edu}

\begin{abstract}
We present observations and analysis of PS1-10bzj, a superluminous supernova (SLSN) discovered in the Pan-STARRS Medium Deep Survey at a redshift $z = 0.650$. Spectroscopically, PS1-10bzj is similar to the hydrogen-poor SLSNe 2005ap and SCP 06F6, though  with a steeper rise and lower peak luminosity ($M_{\rm bol} \simeq -21.4$~mag) than previous events. We construct a bolometric light curve, and show that while PS1-10bzj's energetics were less extreme than previous events, its luminosity still cannot be explained by radioactive nickel decay alone. We explore both a magnetar spin-down and circumstellar interaction scenario and find that either can fit the data. PS1-10bzj is located in the Extended Chandra Deep Field South and the host galaxy is imaged in a number of surveys, including with the {\it Hubble Space Telescope}. The host is a compact dwarf galaxy ($M_B \approx -18$ mag, diameter $\lesssim 800$ pc), with a low stellar mass ($M_* \approx 2.4 \times 10^7 $~M$_{\odot}$), young stellar population ($\tau_* \approx 5$ Myr), and a star formation rate of $\sim 2-3$~M$_{\odot}$ yr$^{-1}$. The specific star formation rate is the highest seen in an SLSN host so far ($\sim 100$ Gyr$^{-1}$). We detect the [\ion{O}{3}] $\lambda 4363$ line, and find a low metallicity: 12+(O/H) = $7.8 \pm 0.2$ ($\simeq 0.1 Z_{\odot}$). Together, this indicates that at least some of the progenitors of SLSNe come from young, low-metallicity populations.
\end{abstract}


\section{Introduction}

The discovery of ``superluminous'' supernovae (SLSNe), with peak luminosities $30-100$ times brighter than normal supernovae and radiated energies $\gtrsim 10^{51}$~erg, is one of the most unexpected results from blank-field time-domain surveys like Pan-STARRS (PS1), Palomar Transient Factory (PTF) and the Catalina Real-Time Transient Survey (CRTS). Several distinct subclasses has been identified, indicating different mechanisms to power the extreme luminosities. Some SLSNe can be classified as Type IIn, likely powered by interaction with a dense, H-rich circumstellar medium \citep[e.g.][]{ock+07,slf+07, scs+10, rfg+11, mbt+13}. The superluminous SN\,2007bi was proposed to be a pair-instability explosion and so ultimately powered by radioactivity \citep{gmo+09}, though this claim is controversial \citep{dhw+12}. The recently discovered SLSN PS1-10afx \citep{cbr+13} does not resemble any previous SLSNe and may define another class of objects.

A third subclass of hydrogen-poor SLSNe similar to the transients SN\,2005ap \citep{qaw+07} and SCP 06F6 \citep{bdt+09} have also been identified, characterized by  blue spectra with a few broad features not matching any standard supernova class \citep{qaw+07, qkk+11, bdt+09, psb+10, ccs+11, lcd+12, bcl+12}. While the associated energetics, ejecta masses and host environments point toward the explosion of a young, massive star, the ultimate energy source remains unknown for these objects. Like the H-rich SLSNe, models based on circumstellar interaction has been proposed \citep{ci11, gb12, mm12}, but the lack of hydrogen seen in the spectrum requires such interaction to be dominated by intermediate-mass elements. Alternatively, the luminosity could be explained by energy injection from a central engine, such as the spin-down of a newborn magnetar \citep{woo10, kb10, dhw+12}. This class has also been linked to Type Ic SNe through the late-time spectroscopic evolution of a few objects \citep{psb+10, qkk+11}, but the relationship between the classes remains unclear. Exploring the diversity of SLSNe and mapping the distribution of explosion properties will be important in further shedding light on the possible energy sources.

Another clue to the progenitor systems could come from studying the host environments. Of the 10 2005ap-like H-poor SLSNe published prior to this paper, only five have detected host galaxies \citep{nsg+11, lcd+12, bcl+12, csb+13}. The host galaxy of SN\,2010gx is the only one that has been studied in detail so far, and is a dwarf galaxy with a low metallicity ($Z = 0.06 Z_{\odot}$), leading to speculation of whether metallicity plays a role in the progenitor channel \citep{csb+13}. Increasing and characterizing the sample of SLSN host galaxies is essential for testing this hypothesis, and constraining the possible progenitors to these extreme explosions.

Here, we present the discovery and analysis of PS1-10bzj, a hydrogen-poor SLSN at $z = 0.650$ from the Pan-STARRS Medium-Deep Survey (PS1/MDS).  We present a comprehensive study of the SN and its host environment. The discovery and observations of PS1-10bzj are described in Section~\ref{sec:obs}. We analyze the properties of the supernova, including temperature evolution, bolometric light curve, possible models, and spectral modeling, in Section~\ref{sec:sn_prop}. Since PS1-10bzj is located in the Extended Chandra Deep Field South (ECDF-S), its host galaxy is detected in the GEMS, GaBoDs and MUSYC surveys \citep{rbb+04, tfd+09,cdu+10}, as well as in the PS1 pre-explosion images. The host galaxy properties, including metallicity, star formation rate, stellar mass and population age, are analyzed in Section~\ref{sec:hostgal}. We place this SN in a broader context, comparing it to previous events, and summarize our results in Sections~\ref{sec:disc} and \ref{sec:conc}. All calculations in this paper assume a $\Lambda$CDM cosmology with $H_0 = 70$~km~s$^{-1}$~Mpc$^{-1}$, $\Omega_{\rm M} = 0.27$ and $\Omega_{\Lambda} = 0.73$ \citep{ksd+11}.

\section{Observations}
\label{sec:obs}

\subsection{PS1 Survey Summary}

The PS1 telescope on Haleakala is a high-etendue wide-field survey
instrument with a 1.8-m diameter primary mirror and a $3.3^\circ$
diameter field of view imaged by an array of sixty $4800\times 4800$
pixel detectors, with a pixel scale of $0.258''$
\citep{PS1_system,PS1_GPCA}.  The observations are obtained through
five broad-band filters (\grizy), with some differences relative to
the Sloan Digital Sky Survey (SDSS); the \gps\ filter extends $200$
\AA\ redward of $g_{\rm SDSS}$ to achieve greater sensitivity and
lower systematics for photometric redshifts, and the \zps\ filter
terminates at $9300$ \AA, unlike $z_{\rm SDSS}$ which is defined by
the detector response \citep{tsl+12}.  PS1 photometry is in the
``natural'' PS1 magnitude system, $m=-2.5{\rm log}(F_\nu) +m'$, with a single
zero-point adjustment ($m'$) in each band to conform to the AB
magnitude scale, determined
with PS1 observations of HST Calspec spectrophotometric standards
\citep{bdc01}.  Magnitudes are interpreted as being at the top of
the atmosphere, with 1.2 airmasses of atmospheric attenuation included
in the system response function \citep{tsl+12}.

The PS1 MDS consists of 10 fields (each with a
single PS1 imager footprint) observed in \gps\rps\ips\zps with a typical cadence of 3~d in each filter, to a $5\sigma$ depth of $\sim 23.3$ mag; \yps is observed near full moon with a typical depth of $\sim 21.7$ mag.
The standard reduction, astrometric solution, and stacking of the
nightly images is done by the Pan-STARRS1 IPP system \citep{PS1_IPP, PS1_astrometry} on a computer cluster at the Maui High Performance Computer Center. The nightly Medium Deep stacks are transferred
to the Harvard FAS Research Computing cluster, where they are
processed through a frame subtraction analysis using the {\tt photpipe}
pipeline developed for the SuperMACHO and ESSENCE surveys \citep{rsb+05, gsc+07, mpr+07}, which was further
improved in order to  increase the accuracy (A.~Rest et al, in preparation, D.~Scolnic
et al., in preparation). The discovery and data presented here are from
the {\tt photpipe} analysis.

\subsection{Photometry}

PS1-10bzj was discovered in PS1 MD02 data on the rise on UT 2010 Dec 16, at coordinates RA=\ra{03}{31}{39.826}, Dec=\dec{-27}{47}{42.17} (J2000). Spectroscopic follow-up confirmed it to be at redshift $z=0.650$ from host galaxy emission lines, placing the peak observed absolute magnitude at $\lesssim -21$~mag, thus classifying it as ``superluminous''. The transient was detected in \gps\rps\ips\zps\ until PS1 stopped observing the field in early 2011 February. All photometry is listed in Table~\ref{tab:phot}, and is corrected for foreground extinction with $E(B-V) = 0.008$ mag \citep{sf11}. When PS1 resumed observing this field in late 2011 September, PS1-10bzj had faded below the detection limit of $\sim 23.5$~mag.

In addition to the PS1 photometry, $griz$ images were obtained along with spectroscopic observations with the Low Dispersion Survey Spectrograph (LDSS3) on the 6.5-m Magellan-Clay telescope, the Inamori-Magellan Areal Camera and Spectrograph (IMACS; \citealt{dhb+06}) on the 6.5-m Magellan-Baade telescope, and the Gemini Multi-Object Spectrograph (GMOS; \citealt{hja+04}) on the 8-m Gemini-South telescope, allowing us to extend the light curve until early 2011 April. These images were reduced using standard routines in IRAF\footnote{IRAF is
  distributed by the National Optical Astronomy Observatory,
    which is operated by the Association of Universities for Research
    in Astronomy, Inc., under cooperative agreement with the National
    Science Foundation.}, and transient flux was determined by subtracting the PS1 pre-explosion template images using ISIS \citep{al98} to correct for galaxy contamination, and measuring the flux in the difference image using aperture photometry. For the LDSS3 and IMACS images, calibrations were obtained either from observations of standard fields on the same night, or from the PS1 catalogs of stars in the field of PS1-10bzj corrected to the SDSS system by the relations in \citet{tsl+12}. In the case of Gemini, archival zeropoints were used for calibration, after verifying that they produce consistent results with the PS1 catalog.

In general the slight difference between the PS1 filter set and $griz$ would not introduce any significant errors. At the particular redshift of PS1-10bzj, however, the [\ion{O}{3}] $\lambda$5007 galaxy emission line is located at the edge between the $i$ and $z$ bands, contributing primarily to the \zps-filter in the PS1 photometric system, but to the $i$-band filter in the SDSS system used at Magellan and Gemini. This line contributes a substantial fraction of the galaxy flux (see Section~\ref{sec:hostgal}). Therefore, non-PS1 $i$ and $z$ fluxes were either determined by subtracting the galaxy templates taken at Gemini and Magellan after the supernova had faded (Section~\ref{sec:hgal_phot}), or corrected according to numerical subtraction.

Figure~\ref{fig:obslc} shows the observed light curves. Since PS1 was observing this field prior to the detection, we are able to constrain the rise time, particularly in \ips-band, where PS1-10bzj brightened by $ >1.2 $ mag in 12 days in the observed frame, corresponding to just 7 days in the rest frame. We also note that the later peak times in the redder bands indicate temperature evolution. Since the best-fit peak is different in different bands, we fit a low-order polynomial to our constructed bolometric light curve (Section~\ref{sec:bol}) to determine the time of maximum light as UT 2011 January 02.65 (MJD 55563.65) $\pm 2$~d. All phases listed are in rest-frame days with respect to this zeropoint.

\begin{figure} 
 \includegraphics{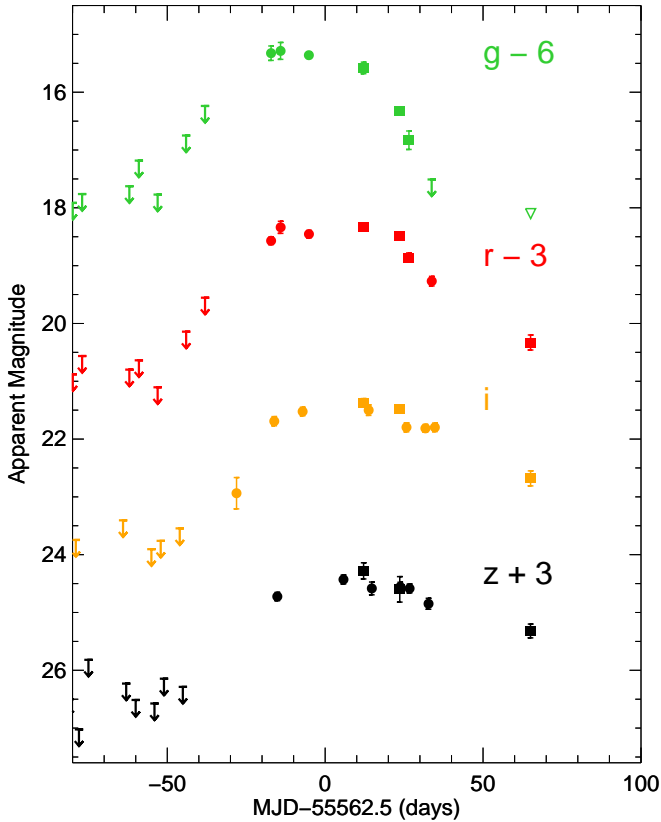}
  \caption{Observed light curve of PS1-10bzj. Time is shown in observer frame, relative to 2011 January 1 (MJD 55562.5). Circles show PS1 photometry, while the squares are photometry obtained with Magellan or Gemini. The arrows and triangles similarly show $3\sigma$ upper limits from PS1 and non-PS1 photometry, respectively. We note the rapid rise time and the faster fall-off in the bluer bands, which indicates temperature evolution.\label{fig:obslc}}
\end{figure}

\subsection{Spectroscopy}

We obtained four epochs of spectroscopy of PS1-10bzj.  Details are
given in Table~\ref{tab:spec}.  Our initial
spectra were taken on 2011 January 18.2 using LDSS3 on the 6.5-m Magellan Clay telescope.
Subsequent observations were obtained with GMOS on the 8-m Gemini-South telescope \citep{hja+04}.
Continuum and arc lamp exposures were obtained immediately after each
object observation to provide a flat field and wavelength
calibration. Basic two-dimensional image processing tasks were
accomplished using standard tasks in IRAF.  Observations of spectrophotometric standard
stars were obtained on the same night as the LDSS3 data, while
archival observations were used for the GMOS spectra.  Our own IDL
routines were used to apply a flux calibration and correct for telluric
absorption bands.

The LDSS3 observations covered the range $3540-9450$~\AA\ in a single
setup using the VPH-all grating and a 0.75$\arcsec$ 
slit oriented at the parallactic angle.
Although no order-blocking filter was used for the object
observations, this setup exhibits very little second-order light
contamination, which we confirmed from observations of
standard stars taken both with and without a
filter, so we believe the spectral shape to be reliable.

The January 25 and 28 GMOS observations were taken with complementary blue
and red setups, which we will sometimes present as a combined single
spectrum.  The January 25 blue spectra were
taken with the slit oriented at a position angle of 175$\degr$, about
68$\degr$ away from the parallactic angle, so differential light loss \citep{fil82}
makes the blue continuum slope on that date unreliable.
The other GMOS spectra were acquired in red setups either at low
airmass (January 28) or near the parallactic angle (April 2 and 3), so
their spectral slopes are reliable.

Our last GMOS observations on April 2 and 3 were obtained in
nod-and-shuffle mode \citep{gb01}.  An error
resulted in the object being nodded off the slit for half of the April
2 observations.  The exposure time quoted in Table~\ref{tab:spec}
reflects only the on-slit time.  The April 2 and 3 data were combined
into a single spectrum.

The spectra from January 18, 25 and 28 are shown in Figure~\ref{fig:spec1}. The April 2 spectrum is dominated by host galaxy light, and is shown in Section~\ref{sec:hostgal}. All of our spectra show a number of narrow emission lines originating in the host galaxy, allowing us to determine a consistent redshift of $z = 0.650$ for PS1-10bzj.

\subsection{Host Galaxy Photometry}
\label{sec:hgal_phot}
The host galaxy of PS1-10bzj is detected in the PS1 pre-explosion stacked images in \gps\rps\ips\zps. In addition, we obtained deep host images with Gemini-S/GMOS, Magellan-Clay/LDSS3 and Magellan-Baade/IMACS in $griz$ after the supernova had faded. Deep infrared imaging in $J$ and $K$ with the FourStar Infrared Camera on Magellan-Baade \citep{pbb+08} yielded only upper limits. Table~\ref{tab:host} lists all galaxy photometry. 

PS1-10bzj is located in the ECDF-S and so photometry from a number of other surveys is also available. From the GEMS survey, there is {\it Hubble Space Telescope} ({\it HST}) imaging with the Advanced Camera for Surveys (ACS) in F606W and F850LP \citep{rbb+04}, and we retrieved the reduced images from the Mikulski Archive for Space Telescopes. From the GaBoDs survey \citep{tfd+09}, there are detections in $U_{38}UBVRIz'$ (and non-detections in $JHK$). In addition, the MUSYC survey \citep{cdu+10} provides imaging in 18 narrow-band filters. This field is also covered by the Spitzer Infrared Array Camera (IRAC) SIMPLE survey \citep{dld+11}, but the galaxy is not detected in their catalog. Archival photometry of the host from the catalogs of these surveys is also included in Table~\ref{tab:host}. 

The various photometric measurements agree well in the bluer filters, but in the different $i$- and $z$-bands there is considerable discrepancy (e.g. the \ips measurement is $\sim0.6$~mag fainter than the GMOS $i$-band, while the \zps measurement is $\sim0.9$~mag brighter than the corresponding $z$ filter). This is explained by the redshifted [\ion{O}{3}] $\lambda 5007$ emission line, located near the edge between $i$ and $z$. The flux we measure in this line from the spectra (Section~\ref{sec:hostgal}) is consistent with the differences in photometry. The effect of this line is also clearly seen in the narrow-band photometry in the IA827 filter.

\section{Supernova Properties}
\label{sec:sn_prop}

\subsection{Spectroscopic and Light Curve Comparisons}
\label{sec:sn_basic}

Given the redshift of $z = 0.650$, we find that PS1-10bzj reached a peak absolute magnitude of $-21.17 \pm 0.15$~mag in \gps. This is luminous enough to be classified as ``superluminous'' according to the definition suggested in \citet{gal12}. Figure~\ref{fig:spec1} shows our spectra of PS1-10bzj, compared to hydrogen-poor SLSNe PS1-10ky \citep{ccs+11} and SN\,2010gx \citep{psb+10, qkk+11}. Our first spectrum shows a blue continuum with broad UV features that are characteristic of the class of hydrogen-poor SLSNe \citep{qkk+11, ccs+11}. These features are also visible in the blue GMOS spectrum taken 7 rest-frame days later. In the red GMOS spectrum on day 16, a number of broad, low amplitude features have also developed, similar to the features seen in SN\,2010gx at late time. The combination of its luminosity and spectral features unambiguously establishes PS1-10bzj as another member of the class of 2005ap-like, hydrogen-poor SLSNe.

Figure~\ref{fig:complc} shows the light curve of PS1-10bzj in absolute magnitude versus rest frame phase compared to a few other hydrogen-poor SLSNe: PS1-10ky and PS1-10awh \citep{ccs+11}, SN\,2010gx \citep{psb+10, qkk+11} and PTF09cnd \citep{qkk+11}. We do not carry out detailed $k$-corrections due to the uncertainties in the spectral energy distributions (SEDs), but have picked bands at similar effective wavelengths as indicated on the plots to facilitate comparisons. With a fast rise time and slower decline, PS1-10bzj does not show the clearly symmetric light curve behavior seen in previous hydrogen-poor SLSNe \citep{qkk+11}, though we note that the rise time is less well constrained in the bluer bands due to shallower limits prior to detection. In general, the light curve of PS1-10bzj exhibits similar timescales to SN\,2010gx and PS1-10ky, but has a flatter peak and is fainter overall.

\begin{figure} 
 \includegraphics{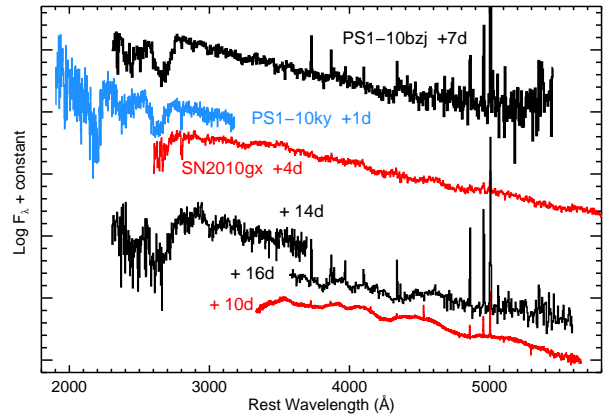}
  \caption{Spectra of PS1-10bzj (black), compared to spectra of other hydrogen-poor SLSNe PS1-10ky (blue; \citealt{ccs+11}) and SN\,2010gx (red; \citealt{psb+10}). The blue continuum and broad UV features are common to hydrogen-poor SLSNe. By the Day 16 spectrum, PS1-10bzj had also developed a number of features in the optical, similar to those seen in SN\,2010gx \citep{psb+10}. See Section~\ref{sec:synow} for modeling and identification of the features.  
\label{fig:spec1} }
\end{figure}

\begin{figure}
   \includegraphics{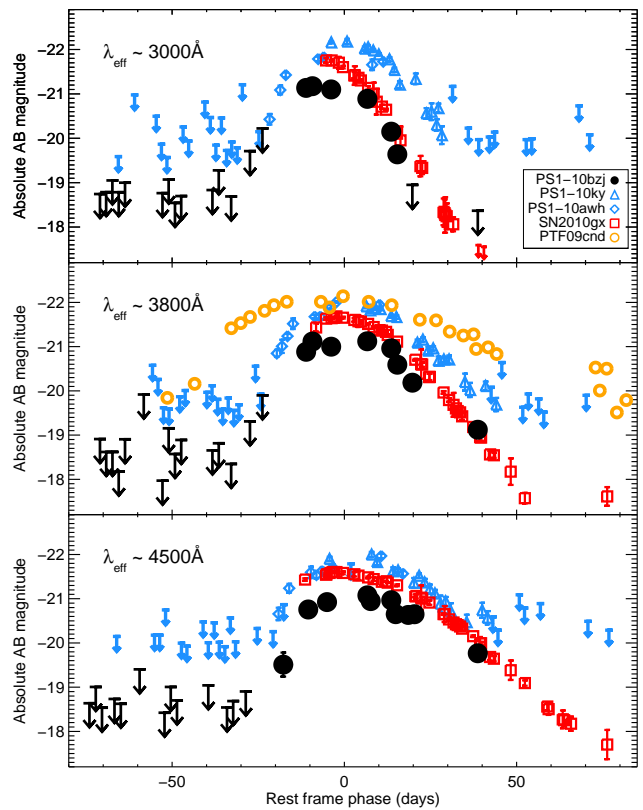}
  \caption{Absolute magnitude light curve of PS1-10bzj (solid black circles and black arrows) compared to other hydrogen-poor SLSNe at similar rest-frame wavelengths, showing \gps ($\sim 3000$~\AA\ rest-frame, top), \rps ($\sim 3800$~\AA\ rest-frame, middle), and \ips ($\sim 4600$~\AA\ rest-frame, bottom). Blue diamonds and triangles show PS1-10awh and PS1-10ky \rps, \ips and \zps respectively \citep{ccs+11}, red squares show SN\,2010gx in $u$, $g$ and $r$ \citep{psb+10, qkk+11}, and yellow open circles show PTF09cnd at $\sim 3600$~\AA\ \citep{qkk+11}.
\label{fig:complc}}
\end{figure}

\subsection{Temperature Evolution and Bolometric Light Curve}
\label{sec:bol}
We determine blackbody temperatures by fitting Planck functions to the broadband photometry, using a $\chi^2$-minimization procedure. For the PS1 photometry, where different bands are observed on consecutive rather than the same night, we first interpolate the photometry to a common time by fitting a low-order polynomial to the nearby light curve points. The SED fits are shown in Figure~\ref{fig:bbfit}, with the model temperatures and radii indicated. These numbers should be interpreted with some caution, as especially at later times the spectrum clearly deviates from that of a blackbody. In addition, by the time of the first spectrum the broad UV absorption features is clearly affecting the $g$-band flux, so that the temperature inferred from photometry is lower than that found by fitting to the spectrum.

The resulting blackbody temperatures and radii from all the fits to the photometry is shown in the top two panels of Figure~\ref{fig:bb_tr_lum}. Prior to peak, we can only place a lower limit on the temperature of $\sim 15,000$~K, since the peak of the blackbody curve is bluewards of the observed photometry and we are essentially fitting the Rayleigh-Jeans tail. After the peak, we find a clear decline in temperature and increase in radius. The best-fit straight line to the estimated blackbody radius (Figure~\ref{fig:bb_tr_lum}) corresponds to an expansion velocity of $11,000 \pm 2000$~\kms, in good agreement with velocities derived from spectroscopic features (Section~\ref{sec:synow}). We note that dividing the estimated radius at peak by the velocity gives a timescale of $\sim 23$ days, consistent with the observed photometric rise.

To construct a bolometric light curve, we first sum the observed flux by trapezoidal integration, interpolating to the edges of the observed bands. Since this only takes into account the flux in the observed wavelength range, it should be considered a strict lower limit of the total radiated power. Integrating this luminosity over the time period we observed the SN indicates a lower limit of the radiated energy $E_{\rm rad} \gtrsim (2.4 \pm 0.5) \times 10^{50}~\mbox{erg}$. The resulting light curve is shown as open red circles in the bottom panel of Figure~\ref{fig:bb_tr_lum}.

To improve this estimate, following \citet{ccs+11} we also include a blackbody tail redwards of the observed bands, using the temperatures determined by our blackbody fits. The resulting pseudo-bolometric light curve is shown as the black, filled circles the bottom panel of Figure~\ref{fig:bb_tr_lum}. We also include the early \ips detection, assuming the same bolometric correction as the next light curve point. PS1-10bzj reached a peak bolometric magnitude $M_{\rm bol} = -21.4 \pm 0.2$~mag, and an estimated total radiated energy of $E_{\rm rad} \gtrsim (3.5 \pm 0.6) \times 10^{50}~\mbox{ergs}$. As expected from the light curve, this is significantly less luminous than previous events - for example, PS1-10awh and PS1-10ky reached peak bolometric magnitudes of $-22.2$~mag and $-22.5$~mag respectively \citep{ccs+11}. While the spectroscopic features clearly identify PS1-10bzj as a member of the same class of objects, it is one of the least luminous hydrogen-poor SLSNe discovered to date. 

\begin{figure}
  \includegraphics{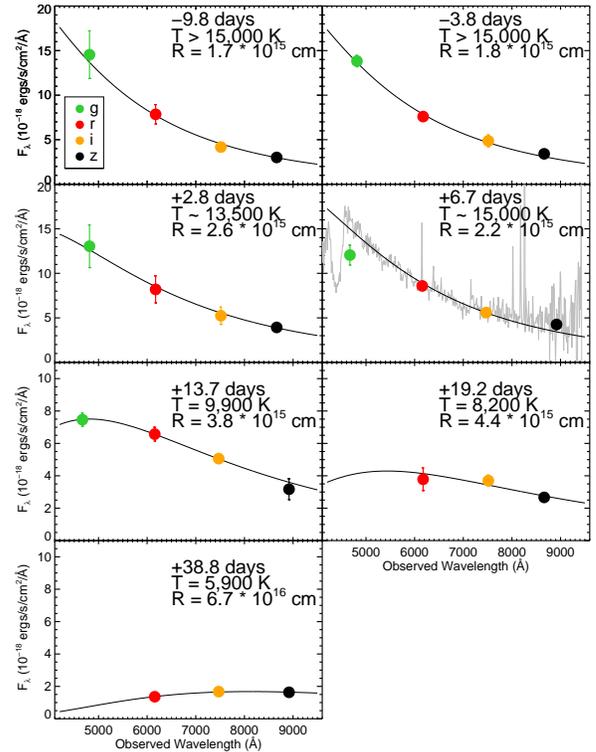}
  \caption{Spectral energy distribution fits to the photometry. The best-fit blackbody temperatures and radii are indicated in the individual panels. The concurrent spectrum is shown with the photometry in the fourth panel; we adopt the temperature derived from the spectrum rather than from the photometry for this date.
  \label{fig:bbfit}}
\end{figure}

\begin{figure}
  \includegraphics{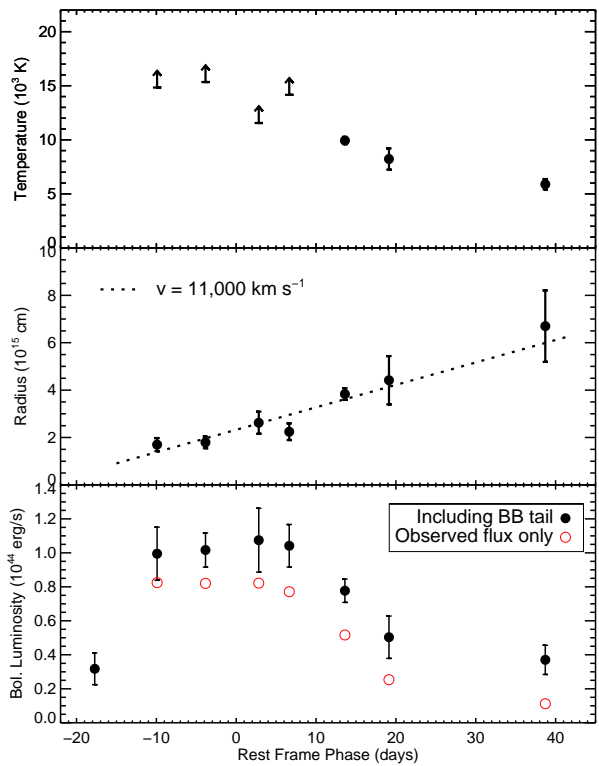}
  \caption{Top panel: temperature evolution of PS1-10bzj, as determined from fitting a blackbody curve to the observed photometry. The uncertainty at early times is largely due to the peak of the blackbody being blueward of our bluest bands; see Figure~\ref{fig:bbfit}. Middle panel: radius of PS1-10bzj, as measured from the same blackbody fit to photometry as the temperature. The best-fit straight line (dashed) corresponds to an expansion velocity of $11,000 \pm 2000$~\kms. Bottom panel: estimated bolometric light curve of PS1-10bzj. The open red circles show observed flux only, while the black filled circles include the observed flux plus a blackbody tail in the red.
\label{fig:bb_tr_lum}}
\end{figure}

\subsection{Line Identifications}
\label{sec:synow}
We used the supernova spectrum synthesis code \texttt{SYNOW} to obtain line identifications and estimates of the expansion velocities, including manual and automated procedures employing the recently updated versions of the software \texttt{SYN++} in combination with \texttt{SYNAPPS}.\footnote{Software was retrieved from https://c3.lbl.gov/es/} The basic assumptions of \texttt{SYNOW} include spherical symmetry, velocity proportional to radius, a sharp photosphere, line formation by resonant scattering treated in the Sobolev approximation, local thermodynamic equilibrium for the level populations, no continuum absorption, pure resonance scattering, and only thermal excitations. Fits are constrained by how we are able to best match absorption minimum profiles, as well as the relative strengths of all the features (see \citealt{bbk+02} for more description of fitting parameters and \citealt{tnm11} for software details).

The fit to the 2011 January 13 spectrum (phase $+7$ d) is shown in the left panel of Figure~\ref{fig:synow}. The photospheric velocity is set at
13,000 \kms, and the temperature to 15,000 K. A maximum cut-off
velocity of 40,000 \kms\ was used for all ions, with the minimum
velocity set to 13,000 \kms. These parameters are
comparable to the ions and associated velocities identified in other
SLSNe (e.g., \citealt{qkk+11,ccs+11}). Two strong features
observed around 2440 and 2650 \AA\ are fit reasonably well with
\ion{Si}{3} and \ion{Mg}{2}, respectively. Introduction of \ion{Fe}{2}
improves the fit around the \ion{Si}{3} line, as seen in the inset. Without \ion{Fe}{2}, the
red wing of the absorption could not be fit with \ion{Si}{3} alone.

Additional weaker features with less certain identifications are also
seen. A sharp cut-off around 3000 \AA\ is likely attributable in part
to \ion{Si}{3}, and we fit absorption features around 4230 and 4490
with \ion{O}{2}. We include in the synthetic spectrum \ion{C}{2},
which is cut off to the blue of \ion{Si}{3}, but is seen in the other
SLSNe and in the later spectrum of this object.

The 2011 January 25 and 28 spectra were combined to fit a single phase $+15$ d
spectrum, shown in the right panel of Figure~\ref{fig:synow}. The photospheric velocity is
set at 11,000 \kms\ and the temperature to 11,500 K. The maximum cut-off
velocity was once again set to 40,000 \kms\ for all ions, which were
fitted with minimum velocities ranging between $11,000$ and $15,000$ \kms.
We observe the same absorption features associated with \ion{Si}{3}, \ion{Mg}{2},
and \ion{C}{2}. Again, including \ion{Fe}{2} substantially improves the fit in this region; the contribution from \ion{Fe}{2} only to the fit is shown in the inset. However, the lack of additional lines elsewhere in the spectrum at this temperature, as well as the noise at the bluest wavelengths, prevent determining the relative strengths accurately. Additional weaker features are fit with \ion{Ca}{2} and \ion{Si}{2}. The \ion{O}{2} seen in the earlier spectrum no longer
appears to be a conspicuous contributor to the spectrum.

Two other SLSNe have shown significant spectral evolution post-peak: in both SN\,2010gx and PTF09cnd \citep{psb+10,qkk+11}, the spectra evolved to look like normal Type Ic SNe at late times. The connection between H-poor SLSNe and SNe Ibc is also suggested by a transient ``W''-shaped feature near 4200 AA seen in early spectra of the well-observed Type Ib SN 2008D, identified as the same feature as seen in SN 2005ap \citep{mlb+09, qaw+07}. Following \citet{qaw+07}, \citet{mlb+09} modeled this feature with a blend of \ion{O}{3}, \ion{N}{3} and \ion{C}{3}; later modeling by \citet{qkk+11} of SN 2005ap and other H-poor SLSNe updated the identification of the ``W''-feature to \ion{O}{2}. The presence of this transient feature in SN 2008D thus provides an additional link between Type Ibc SNe and SLSNe. A basic question is whether the SLSNe are truly distinct objects from normal Type Ic SNe, or whether there is a smooth continuum between the two. With its comparatively low peak luminosity, PS1-10bzj is closer in luminosity to luminous Type Ic SNe like SN\,2010ay, which peaked at $M_r = -20.2$ \citep{ssv+12}, than to the prototype SLSNe 2005ap and SCP 06F6. From this perspective, it is interesting to note that at least over the time we were following it, the spectral features in PS1-10bzj do not look like SN Ic features, including objects like SN\,2010ay.

\begin{figure*}
\plotone{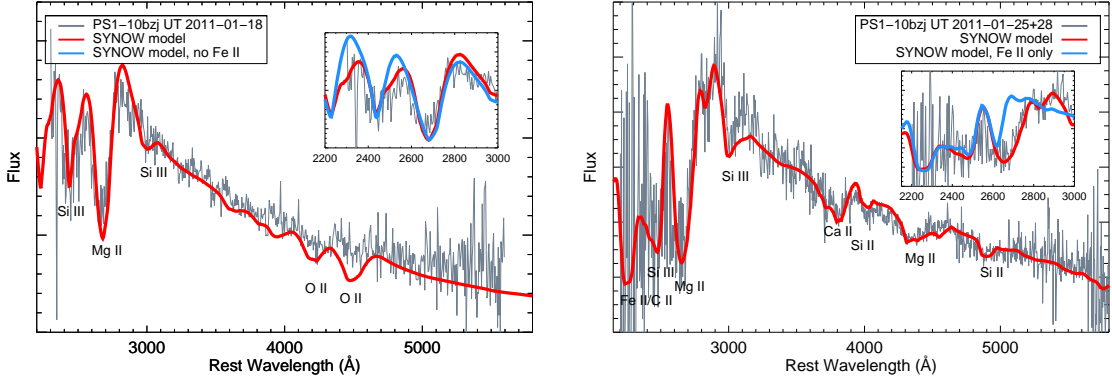}
\caption{SYNOW fits to the spectra of PS1-10bzj. Left: the 2011 January 13 (phase $+7$ d) spectrum. The strong UV features are well fit with \ion{Si}{3} and \ion{Mg}{2}; minor features near 4500\AA\ could be due to \ion{O}{2}. The shape of the \ion{Si}{3} feature is better fit when including \ion{Fe}{2}, as illustrated in the inset, which shows the fit without including this ion.  Right: fit to combined 2011 January 25+28 (phase $+15$ d) spectrum. Minor features in the optical are fit with \ion{Ca}{2}, \ion{Mg}{2} and \ion{Si}{2}. The fit includes a substantial contribution from \ion{Fe}{2} (shown in the inset), but the overlap of these features with the Mg and Si features and lack of features elsewhere prevent us from determining the relative strengths accurately.
\label{fig:synow}}
\end{figure*}

\subsection{Light Curve Model Fits}
The optical luminosity of most canonical Type I SNe (i.e. type Ia, Ib and Ic) is powered by the radioactive decay of $^{56}$Ni, with the shape of the light curve primarily dictated by three parameters: the nickel mass ($M_{\rm Ni}$) which sets the total luminosity, the total kinetic energy ($E_{\rm K}$), and ejecta mass $M_{\rm ej}$, which set the characteristic time of photon diffusion $\tau_c \propto M_{\rm ej}^{3/4} E_{\rm K}^{-1/4} $ and essentially determines the width of the light curve \citep{arn82}. Measurements of the photospheric velocity ($v_{\rm ph}$) from the spectra constrain $\sqrt{E_{\rm K} / M_{\rm ej}}$, so that all three parameters can be determined based on observable quantities. We fit our bolometric light curve of PS1-10bzj using the models of \citet{vbc+08} and \citet{dsg+11}; see Figure~\ref{fig:nickel}. The light curve can be reasonably fit with $M_{\rm Ni} \simeq 6-8~$M$_{\odot}$, with the best-fit model having $M_{\rm Ni} = 7.2~$M$_{\odot}$ and $\tau_c \simeq 19$~d. Using the photospheric velocity derived from the spectrum near peak, $v_{\rm ph} = 13,000$~km~s$^{-1}$, yields $M_{\rm ej} \sim 5-11~$M$_{\odot}$, with $8.5~$M$_{\odot}$ for the best-fit model. Therefore, if PS1-10bzj were powered by radioactive decay, it would require a $^{56}$Ni mass of $\gtrsim 10$ times what is observed in typical Type Ibc or Ic-BL SNe ($0.2 - 0.5~{\rm M}_{\odot}$; \citealt{dsg+11}). In addition, the ejecta would have to be  $75-100 \%$ $^{56}$Ni by mass, a fraction seen in no observed SNe, including proposed pair-instability SNe like SN\,2007bi where the inferred Ni mass was several M$_{\odot}$ \citep{gmo+09}. A composition of $> 75$\% Ni would also result in a large amount of line-blanketing in the rest-frame UV, which is not seen. Radioactive decay, then, is unlikely to be the main contributor to the luminosity. This is consistent with what is found for other hydrogen-poor SLSNe \citep{ccs+11, qkk+11}.

\begin{figure}
 \plotone{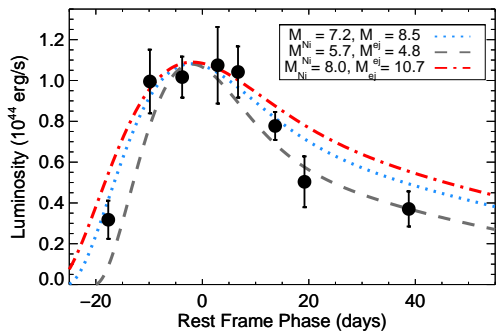}
   \caption{Model of radioactive Nickel decay, following \citet{dsg+11}. The light curve can be reasonably fit with Ni masses in the range $6-8~$M$_{\odot}$, but require the ejecta composition to be $ 75-100\%$ Ni to simultaneously fit the peak luminosity and the light curve width. 
\label{fig:nickel}}
\end{figure}

Since nickel decay is unlikely, other explanations have been proposed for the extreme luminosity of SLSNe. One possibility is energy injection by a central engine, such as the spin-down of a newborn magnetar \citep{kb10, woo10}. We fit our bolometric light curve with the model of \citet{kb10}, following the procedure outlined in \citet{ccs+11}. Our assumptions include magnetic dipole spin-down, an opacity of $\kappa = 0.2$~cm$^2$~g$^{-1}$, and a supernova energy of $10^{51}$~erg; we vary the ejecta mass, the magnetar spin ($p$), and the magnetic field ($B$). With these assumptions, Figure~\ref{fig:magnetar} shows a best-fit model, with $M_{\rm ej} = 2$~M$_{\odot}$, $B = 3.5 \times 10^{14}$~G, and $p = 4$~ms; however we find that the light curve can be reasonably fit within the uncertainties with ejecta masses in the range $M_{\rm ej} \sim 1-6~$M$_{\odot}$.  An ejecta mass lower than 1~M$_{\odot}$ predicts a light curve that is too narrow, while ejecta masses greater than $\sim 6$~M$_{\odot}$ require initial spins faster than the maximum (breakup) spin of $\sim 1$~ms to match the timescales. Within this range, however, parameters can be chosen to fit the light curve equally well within the uncertainties. We note that the inferred ejecta masses are similar to what is seen in normal Type Ibc SNe \citep{dsg+11}.

The magnetar model predicts that the ejecta will be swept up into a dense shell, which then expands at a constant velocity. For the range of models that fit the light curve, those with higher spin periods (and so a lower total energy, as the magnetar contribution scales as $p^{-2}$) generally have lower inferred velocities, in better agreement with the velocities inferred by the spectra. The 2~M$_{\odot}$ model shown in Figure~\ref{fig:magnetar} has the swept-up shell expanding at 11,000~km~s$^{-1}$, in good agreement with the observed velocities. The predicted temperatures for this model also match the observed temperatures within the errors. A simple magnetar model thus provides a reasonable fit to the observed properties of PS1-10bzj. One caveat is that our modeling of the spectra support a slightly declining, rather than constant photospheric velocity. The rapid evolution of the spectrum is also challenging to explain in the context of this model.

\begin{figure}
 \plotone{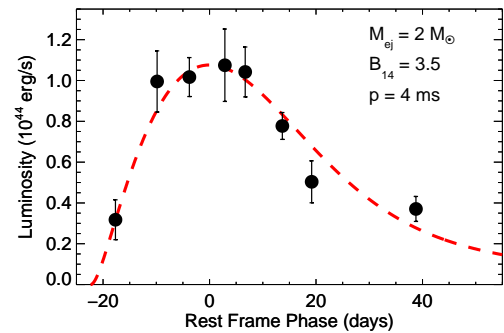}
   \caption{Magnetar model fit to the bolometric light curve, with $M_{\rm ej} = 2$~M$_{\odot}$, $B = 3.5 \times 10^{14}$~G, and $p = 4$~ms. While the observed light curve can be fit with a range of parameters within the uncertainties, the model shown also predicts a velocity and temperature evolution that agrees with the observed data.
\label{fig:magnetar}}
\end{figure}

A third proposed mechanism for powering SLSNe is interaction with opaque, circumstellar material. This leads to efficient conversion of the kinetic energy to radiation, with the resulting lightcurve being due to shock breakout through this opaque wind. This class of models has been applied both to superluminous SNe IIn such as SN\,2006gy, and to SN\,2005ap-like objects \citep{sm07, scs+10, ci11, bl11, gb12, cwv12}. The light curve of the hydrogen-poor and superluminous SN\,2006oz, in particular, showed a ``dip'' feature on the rise that has been interpreted as a signature of shock breakout \citep{lcd+12, mm12}. 

We can use the observed properties of PS1-10bzj and the analytical relations of \citet{ci11} to estimate the physical conditions required in the interaction scenario. Assuming a wind density profile $\rho_w = D r^{-2}$, as expected from a steady wind, so that $D = \dot{M}/4\pi v_w \equiv 5 \times 10^{16} D_*$ in cgs units. The rise-time can be roughly equated to the diffusion time, $t_d = 6.6 \kappa D_* {\rm ~d}$, where $\kappa$ is the opacity in units of $0.34~{\rm cm}^{2}{\rm~g}^{-1}$. We use $\kappa = 0.59$, as expected for an ionized He-rich wind. Taking the rise-time of PS1-10bzj to be $\sim 20$~days, we find that $D_* \simeq 5.1$, which gives a total required wind mass of $\sim 3.5$~M$_{\odot}$, using the radius at peak to be $\sim 2.2 \times 10^{15}{\rm~cm}$. Using $E_{\rm rad} \approx 3.5 \times 10^{50} {\rm~erg}$, we find that the associated supernova energy is $2.2 \times 10^{51} (M_{\rm ej}/10 {\rm M}_{\odot})^{1/2}{\rm~erg~} $, and corresponding diffusion radius $R_d \approx 1 \times 10^{15}{\rm ~cm}$, assuming an ejecta mass of 10~M$_{\odot}$ and using Equations~(5) and (3) in \citet{ci11} respectively. The predicted velocity of the photosphere, using Equation~3 in \citet{ccs+11}, is 11,800~\kms, in good agreement with the observed velocities. Thus, this model can also reproduce the basic observed properties, but require a wind mass of several M$_{\odot}$ of hydrogen-poor material.

Recently, \citet{gb12} have shown that the simple treatment in \citet{ci11} is not appropriate in regimes where the wind radius is comparable to the diffusion radius. Instead, they carried out hydrodynamical simulations of supernovae exploding into dense circumstellar material, successfully matching the light curves of SN\,2005ap, SN\,2006gy and SN\,2010gx. Given the similarities between PS1-10bzj and SN\,2010gx, it seems plausible that its light curve could also be fit by a more sophisticated shock breakout model, though calculating such a model is outside the scope of this paper. We note that the \citet{gb12} model for SN\,2010gx requires an even larger total wind mass ($M_w \simeq 16$~M$_{\odot}$) than our estimate for PS1-10bzj based on the simple \citet{ci11} relations, so an extreme mass loss episode would likely still be required. 

A simple interaction model, then, can also explain the observed data, but requires a mass-loss rate of $\sim 3$~M$_{\odot}{\rm ~yr}^{-1}$ in the last year before explosion, assuming a wind velocity of 1,000~\kms (as seen in Wolf-Rayet stars; e.g. \citealt{nl00}). In addition, the lack of hydrogen and helium seen in the spectra would require this circumstellar material to be primarily composed of intermediate-mass elements. One might also expect to see intermediate-width lines in the spectra if the primary energy source is interaction, but this has not been seen in any of the H-poor SLSNe, including PS1-10bzj. A detailed radiative transfer model is necessary to see if this scenario can reproduce the spectra as well as the light curves.

\section{Host Galaxy Properties}
\label{sec:hostgal}
In addition to studying the SN explosion itself, additional clues to the nature of the progenitors come from studying the host environments of the SLSNe. In the case of PS1-10bzj there is a wealth of data on the host galaxy, allowing for a detailed study.

\subsection{Luminosity and Size}
The absolute magnitude of the host is $M_B = -18.0$~mag, corrected for cosmological expansion and foreground extinction. This is similar to what has been seen for other SLSN hosts, which seem to show a preference for low-luminosity galaxies \citep{nsg+11, csb+13}. In terms of the luminosity function at $z \sim 0.7$ \citep{itz+05,wfk+06}, this corresponds to a $0.05 L_*$ galaxy.

The host galaxy of PS1-10bzj is unresolved in all our ground-based images (with seeing down to $\sim 0.6$\arcsec). Since the field was covered by the GEMS survey, we also have available {\it HST}/ACS images in F606W and F850LP \citep{rbb+04}, shown in Figure~\ref{fig:hst_host}. Even in these images, the host is not obviously resolved with a FWHM of $\sim 0.12\arcsec$ (whereas the mean FWHM of stars in the images is $\sim 0.10\arcsec$).At $z = 0.650$, this corresponds to an upper limit on the galaxy diameter of $\lesssim 800~{\rm pc}$. We note that it is possible that what we see in the {\it HST} images is only one bright knot of star formation, and the galaxy itself could be more extended. Nevertheless, the combination of luminosity and size establishes the host as a compact dwarf galaxy.

\begin{figure}
\begin{center}
 \includegraphics[width=3.5in]{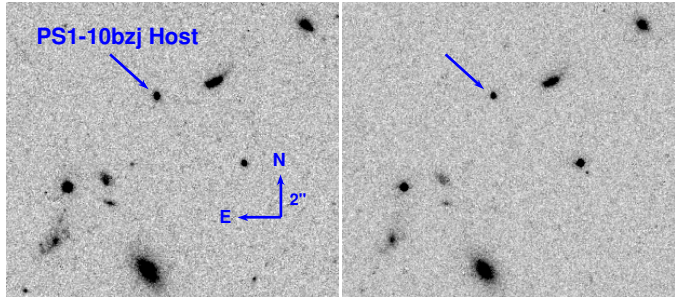}
  \caption{{\it HST}/ACS images of the field around PS1-10bzj, in filters F606W (left) and F850lp (right). The host, marked with the blue arrow, is remarkably compact.
  \label{fig:hst_host}}
\end{center}
\end{figure}

\subsection{Stellar Mass and Population Age}
To determine the stellar mass ($M_*$) and population age ($\tau_*$) of
the host we fit the SED with the
\citet{mar05} evolutionary stellar population synthesis models, using
a Salpeter initial mass function and a red horizontal branch
morphology. Since the model only accounts for the continuum emission, and the flux in the host emission lines is substantial, we restrict our fit to the MUSYC narrow-band filters without significant emission lines. The fit to the SED is shown in Figure~\ref{fig:host}. We find that the host SED is well fit with a stellar population age of $\tau_* \approx 5$ Myr, yielding a stellar mass of $ M_* \approx 2.4 \times 10^7$~M$_{\odot}$. This assumes $A_V = 0$, measured from the Balmer decrement (Section~~\ref{sec:met}).

Another estimate of the stellar population age comes from the
H$\beta$ equivalent width (EW).  While we do not have a galaxy-only
spectrum, the 2011 April 3 Gemini spectrum (Figure~\ref{fig:emlines})
does not show any broad supernova features and is dominated by galaxy
light ($\gtrsim 50\%$; estimated from pre-explosion galaxy photometry).  It can
therefore be used to determine a lower limit on the H$\beta$ EW,
which we find to be $W_r\approx 61$ \AA. This value
yields a young stellar population age of $\lesssim 5$ Myr for a
metallicity $Z =0.2-0.4$ Z$_\odot$, using the fits in \citet{lbk+10}
to the models of \citet{sv98}.  This value is in excellent agreement with
the stellar population age inferred from SED modeling.

\subsection{Metallicity}
\label{sec:met}
While all of our spectra include contributions from both the galaxy and the SN,
they clearly exhibit narrow emission lines originating in
the host galaxy. Figure~\ref{fig:emlines} shows the 2011 April 3 GMOS
spectrum, which is dominated by galaxy light, with the strongest
emission lines marked.  We measure the line fluxes in all of our spectra by fitting Gaussian
profiles (Table~\ref{tab:emlines}).  With the exception of the
[\ion{O}{3}] $\lambda4363$ line, which was only robustly detected in
the 2011 January 28 GMOS spectrum, we use the weighted average of the
three measurements for line diagnostics. Absolute flux calibration is based on the 2011 January 13 LDSS3 spectrum, by scaling synthetic photometry from the spectrum to photometry obtained the same night. The GMOS spectra were then scaled according to the flux in the [\ion{O}{3}] doublet.

None of our spectra cover H$\alpha$, which is located at $1.083~\mu$m at this redshift.  We therefore use the Balmer decrement as
measured from H$\gamma$/H$\beta$ to estimate reddening, assuming a
Case B recombination value of 0.469 \citep{ost89}.  Since our measured
value of $0.48\pm 0.03$ is consistent with no reddening, we conclude
that the host galaxy extinction is minimal.

We detect the auroral [\ion{O}{3}] $\lambda4363$ line in the 2011
January 28 GMOS spectrum at $3.5\sigma$ significance, shown in the inset of Figure~\ref{fig:emlines}.  Assuming an electron density
$n_e = 100$ cm$^{-3}$, we calculate an electron temperature of
$T_e(O^{++})=16,200^{+2,900}_{-1,700}$ K from the ratio of
[\ion{O}{3}] $\lambda4363$ to [\ion{O}{3}] $\lambda\lambda 5007,4959$,
using the IRAF task {\tt temden}.  This result is not sensitive to the
exact choice of density since $T_e$ is insensitive to small changes in
density \citep{kbg+07}; for example, we find identical results when
doubling the assumed electron density to 200 cm$^{-3}$.  Using the
relation $T_e(O^+) = 0.7\times T_e(O^{++})+0.3$ from \citet{sta82}, we
determine $O^+/H$ and $O^{++}/H$ using the relations in \citet{skc06}.
This gives an electron temperature metallicity of $12+\log(\mbox{O/H})
=7.8\pm 0.2$. This translates to $Z = 0.13 Z_{\odot}$, using the solar abundance of \citet{ags+09}. This low abundance is consistent with the inferred young stellar population age and low stellar mass.

For comparison, we also estimate the oxygen abundance using the
$R_{23}$ diagnostic with the calibration of \citet{kk04}.  We
measure $R_{23}\equiv (\mbox{[\ion{O}{3}]} \lambda\lambda 5007,4959 +
\mbox{[\ion{O}{2}]} \lambda 3727)/\mbox{H}\beta=9.25\pm 0.36$, and an
ionization parameter of $y\equiv \log\left([\mbox{\ion{O}{3}}] 
\lambda\lambda 5007,4959 / \mbox{[\ion{O}{2}]} \lambda 3727 \right)
=0.95\pm 0.03$.  Using the iterative scheme in \citet{kk04}, and
assuming the lower metallicity branch based on the presence of the
[\ion{O}{3}] $\lambda4363$ line, this method gives a metallicity
$12+\log(\mbox{O/H})\approx 8.3$. This is 0.5~dex higher than the result from the direct method, but we note that this discrepancy is not unusual; theoretical strong-line indicators are known to be offset from the direct method, with the difference being larger at the lower-metallicity end \citep{bgk+09}. A similar discrepancy is seen in the host galaxy of SN\,2010gx, where \citet{sps+11} found $12 + \log({\rm O/H}) = 8.36$ using the \citet{kk04} calibration, while the direct method yields $12+\log({\rm O/H}) = 7.46$ \citep{csb+13}. Still, strong-line metallicity indicators provide a useful basis for comparison, since direct metallicity indicators are otherwise mostly only available for low-redshift samples.

\subsection{Star Formation Rate}
We estimate the SFR of the host
galaxy from the [\ion{O}{2}] $\lambda3727$ line flux, using the
metallicity-dependent relation in \citet{kgj04}.  Using the
metallicity and ionization parameter we determined from the $R_{23}$
method, we find ${\rm SFR}\approx 2$~M$_\odot$~yr$^{-1}$. Alternatively, since the H$\gamma$/H$\beta$ ratio indicates no extinction, we can use the H$\beta$ flux to predict the H$\alpha$ flux, assuming a ratio ${\rm H}\alpha / {\rm H}\beta = 2.85$ according to case B recombination. Using ${\rm SFR} = 7.9 \times 10^{-42} L_{{\rm H}\alpha} {\rm(erg~s}^{-1})$ \citep{ken98}, we find ${\rm SFR}\approx 4.2$~M$_\odot$~yr$^{-1}$, in reasonable agreement with the [\ion{O}{2}] $\lambda3727$ estimate.

A complementary method to calculating the SFR is to use the galaxy UV
continuum flux.  At this redshift, the $UBg$ filters sample rest-frame
$2300-2900$ \AA, allowing us to use the relation from \citet{ken98}:
${\rm SFR}=1.4\times 10^{-28} L_{\nu}$.  This yields ${\rm SFR}\approx
2-3$ M$_{\odot}$ yr$^{-1}$, also in good agreement with the estimates from emission lines.

Combining the stellar mass with the SFR, we calculate a specific star formation rate (sSFR) of $\sim 100$~Gyr$^{-1}$. This is significantly higher than the $\sim 2.6$ Gyr$^{-1}$ measured in the host of SN 2010gx \citep{csb+13}, and also higher than the $\sim 10$ Gyr$^{-1}$ in the host of PS1-10bam \citep{bcl+12}. The basic picture of a low metallicity, low mass and highly star-forming dwarf galaxy is similar to what has been seen for other SLSNe.

\begin{figure}
\begin{center}
\includegraphics[width=3.3in]{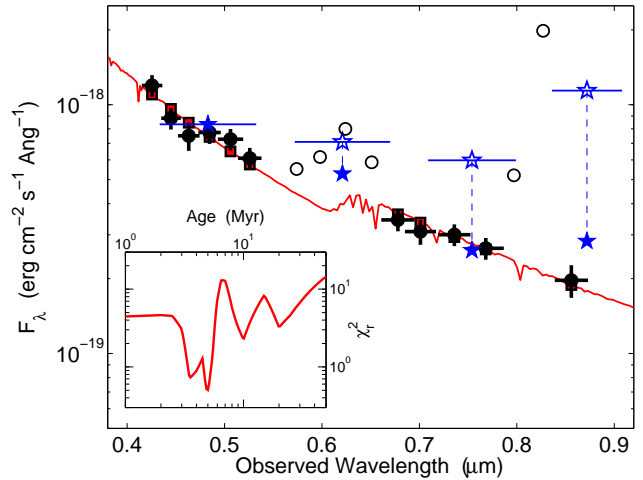}
\caption{Best-fit spectral energy distribution models to the host
galaxy photometry, using $A_V=0$~mag. Only the narrow-band photometry is used in the fit (filled circles), due to the strong emission lines contaminating the broadband filters. Open circles show narrow-band filters which contain emission lines and are therefore not used for the fit. For comparison, the stars show PS1 broad-band photometry, both uncorrected (open) and corrected (filled) for the flux in the emission lines. Additional broad-band filters are largely redundant with the PS1 ones and are not shown.
Our best-fit is a young ($\tau_* \approx 5$ Myr) and low-mass ($M_* \approx 2.4 \times 10^7$~M$_{\odot}$) stellar population; an acceptable fit also exist for a 3.5 Myr population, and slightly worse fits for a 10 or 20 Myr population, as shown in the inset.
\label{fig:host}}
\end{center}
\end{figure}

\begin{figure}
\includegraphics{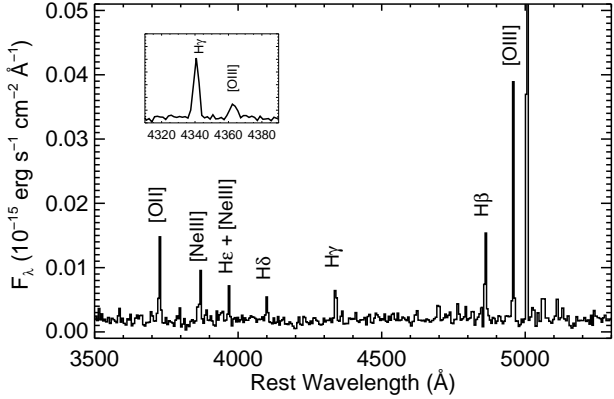}
\caption{2011 April 3 GMOS spectrum of PS1-10bzj, with strong host
galaxy emission lines marked.  While there is some supernova
contribution to the flux in this spectrum, it is dominated by host
galaxy light ($\gtrsim 50\%$). The inset shows the region around the auroral [\ion{O}{3}] $\lambda4363$ line, from the 2011 Jan 28 spectrum, which has somewhat better signal-to-noise ratio.
\label{fig:emlines}}
\end{figure}

\section{Discussion}
\label{sec:disc}

\subsection{The Diversity of SLSNe}

As was shown by \citet{qkk+11}, the hydrogen-poor SLSNe form a spectroscopic class, though with a range of light curve properties. Figure~\ref{fig:peakmag} shows the distribution of peak absolute magnitudes of all published 2005ap-like hydrogen-poor SLSNe, corrected for cosmological expansion by $M  = m - 5\log\left(d_L(z) / 10 {\rm pc}\right) + 2.5\log(1+z)$. Due to the lack of SED information in several objects, we do not carry out full $k$-corrections, but note that where multiband photometry is available the observed peak is at a rest-frame wavelength of $\sim 3000-4000$\AA. SN\,2006oz is not included in this plot, as it was only observed on the rise and so the peak magnitude is not well constrained. Most of the hydrogen-poor SLSNe peak near $M \simeq -22$~mag, with a tail to higher luminosities. The apparent lack of lower-luminosity objects is likely due, at least in part, to the flux-limited surveys (and spectroscopic follow-up) so that there is a bias toward finding brighter objects. PS1-10bzj is both the lowest-luminosity and one of the lowest redshift SLSNe found in PS1/MDS, for example. Recently, \citet{qya+13} found that the distribution of hydrogen-poor SLSNe seems to be narrowly peaked, also when taking the effects of flux-limited selection into account. If so, an event like PS1-10bzj would be intrinsically rarer than the $-22$~mag objects, at the low-luminosity tail of the distribution.

The timescales seen in SLSNe also vary by a factor of several, with rise times varying from $\sim 20$~d in the case of PS1-10bzj, to $\gtrsim 50$~d in PTF09cnd \citep{qkk+11}. If the faster timescales are typical for the lower-luminosity end of the distribution, it may present an additional selection bias against the fainter objects, as the timescales are approaching those of normal SNe and so the objects stand out less amongst the more common normal SNe.

Finding fainter hydrogen-poor SLSNe is particularly interesting because this class is linked to Type Ic SNe through the late time spectroscopic evolution of a few objects. A basic question is whether they are truly distinct populations, or whether there is a smooth transition between the two. There is a luminous tail to the Type Ic distribution: for example SN\,2010ay reached a peak luminosity of $-20.2$~mag, though still had a light curve consistent with being powered by nickel decay and did not show the spectroscopic features typical of SLSNe \citep{ssv+12}. If the dearth of intermediate-luminosity objects represents a real cutoff rather than a selection effect, this places constraints on any proposed mechanism for powering the SLSNe. Such a low-luminosity cutoff is not predicted by theoretical models of SLSNe; for example the magnetar models presented in \citet{kb10} can reproduce a wide range of luminosities and timescales.

We also note that PS1-10bzj only has a few, early light curve points that are brighter than $-21$~mag, but is clearly a spectroscopic member of the class of 2005ap-like hydrogen-poor SLSNe. This suggests that a definition based on a luminosity cut, as was suggested in \citet{gal12}, is artificial and that this class of objects is better defined by spectroscopic features.

\begin{figure}
\includegraphics{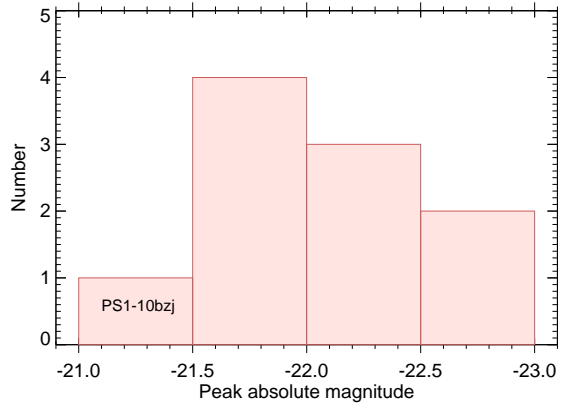}
\caption{Distribution of peak observed absolute magnitudes for the hydrogen-poor SLSNe published to date. SN\,2006oz is not included, as it was only observed on the rise and so only a lower limit $< -21.5$~mag is available.
\label{fig:peakmag} }
\end{figure}

\subsection{The Host Galaxy Environments}

Of the 10 2005ap-like hydrogen-poor SLSNe published prior to this work, only five have detected host galaxies and the upper limits on the undetected ones are $M_r \gtrsim  -18 $~mag \citep{nsg+11}. It has been speculated that this preference for low-luminosity environments is really a preference (and perhaps requirement) for low-metallicity environments \citep{nsg+11, sps+11}. This is supported by the host of SN\,2010gx, the first SLSN host galaxy with a direct metallicity measurement, with $Z = 0.06 Z_{\odot}$ \citep{csb+13}). The low metallicity of 0.13 $Z_{\odot}$ for the host of PS1-10bzj follows the same trend. 

To put these galaxy measurements in context, in Figure~\ref{fig:mass_met} we plot different properties of the two SLSN host galaxies, compared to other galaxy samples. The top left panel shows a mass-metallicity ($M-Z$) plot, with metallicity measured by the $R_{23}$ method from \citet{kk04} to facilitate comparison to different samples, including core-collapse SNe \citep{kk12} and GRB host galaxies \citep{lkb+10,lb10,mkk+08}. We plot here also the host of the superluminous SN\,2007bi, with $R_{23}$ metallicity from \citet{ysv+10}, and mass we estimated from the SDSS photometry of this host. It is worth noting that all three SLSN hosts have very similar $R_{23}$ metallicities. They are all less massive and more metal-poor than the core-collapse SN hosts, but occupy a similar region as the gamma-ray burst (GRB) host galaxies. One important caveat here is that the core-collapse SN hosts are generally at low redshift and contain a mix of hosts from targeted and untargeted surveys, so we would not necessarily expect them to follow the same $M-Z$ relation. 

The remaining three panels plot metallicity as measured by the $T_e$ method, against either luminosity, mass, or the combination of mass and SFR that minimizes scatter in metallicity (the so-called Fundamental Relation or FMR; \citealt{mcm+10,am13}). The host of PS1-10bzj is consistent with each of the nearby relations within its uncertainties, indicating that it is not an unusually metal-poor galaxy given its mass, luminosity and SFR. The similarity to GRB hosts may indicate that the two phenomena happen in similar environments, but the sample sizes here are small. The host of SN\,2010gx stands out as more extreme than the host of PS1-10bzj in terms of metallicity, and falls below the nearby/SDSS relations in each case. The two SLSN hosts are the most separated on the FMR plot, due to the larger sSFR of the PS1-10bzj host. As such, the most striking common factor between the two galaxies is their low metallicities.

 We note that if low metallicity is an important factor in producing this type of SLSN, this may present a challenge for models where the luminosity of the SN is powered by interaction with a dense wind. In particular, the mass loss would be unlikely to be driven by metal-line winds and so a different mass-loss mechanism would be required.

\begin{figure*}
\begin{tabular}{cc}
\includegraphics{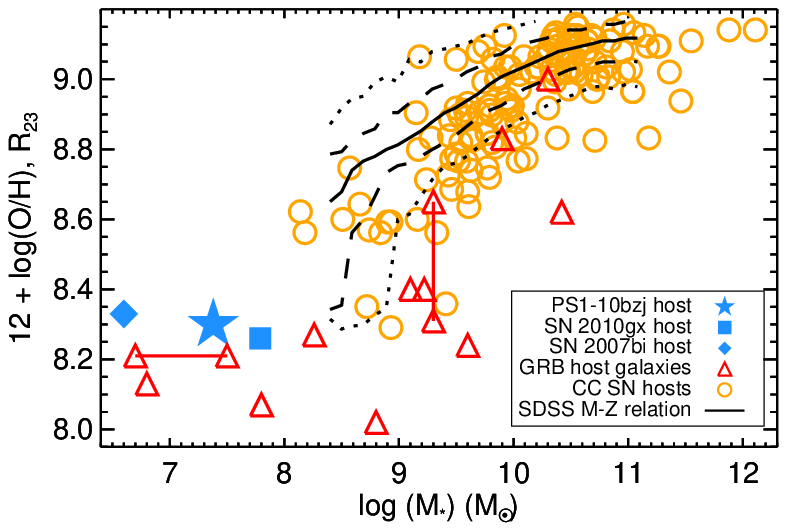} & \includegraphics{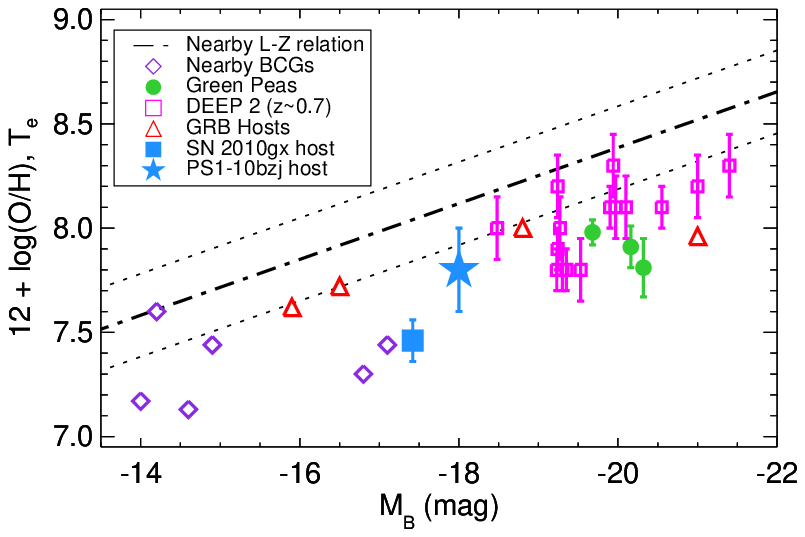} \\
\includegraphics{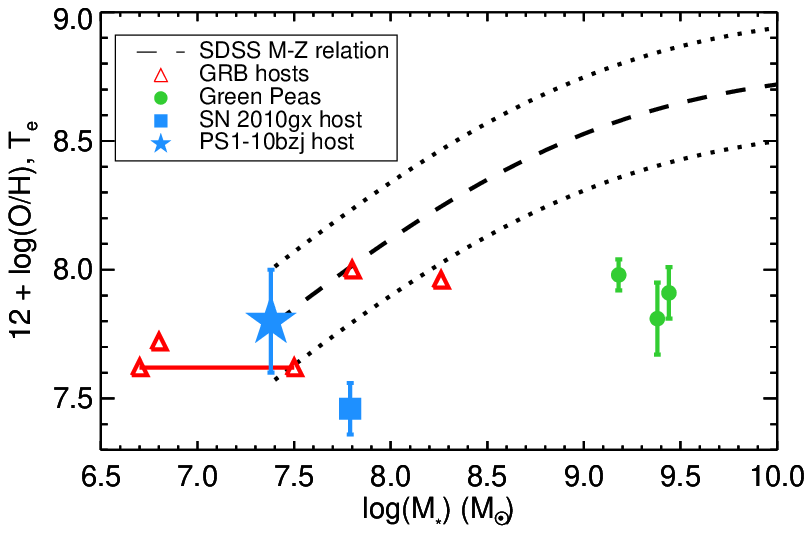} & \includegraphics{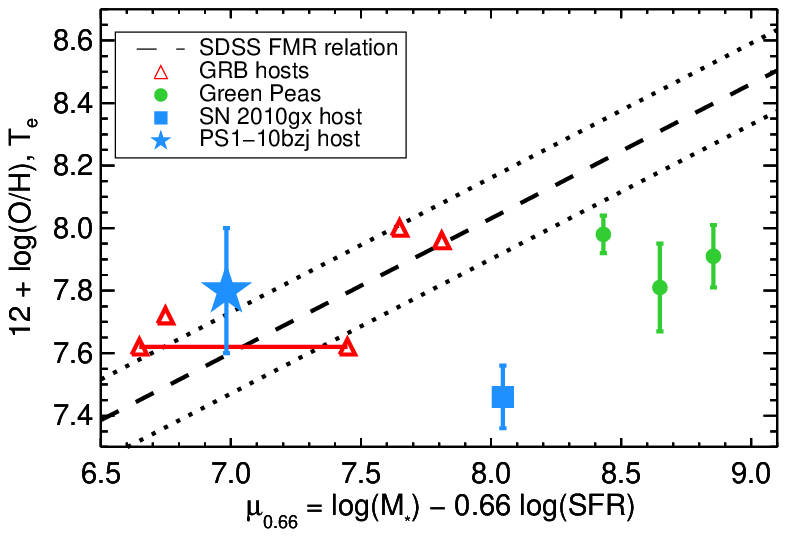} \\
\end{tabular}
\caption{A comparison of the host of PS1-10bzj to other SLSNe and galaxy samples. The blue star and square show the hosts of PS1-10bzj and SN\,2010gx, respectively. The top left panel is a mass-metallicity ($M-Z$) diagram, here plotting metallicity as calculated by the $R_{23}$ method in \citet{kk04}, to facilitate comparison to a broader sample. The black lines show the SDSS $M-Z$ relation \citet{thk+04}, where metallicity has been converted to the KK04 scale using the relations in \citet{ke08}. The orange circles show hosts of core-collapse supernova (any type), with metallicities similarly converted from the \citet{thk+04} scale. Red triangles show GRB hosts; connected points indicate a dual solution for either the mass or the metallicity. The other three plots show metallicity measured by the direct method, comparing to either the nearby luminosity-metallicity relation (\citealt{gpm+09}; top right) the SDSS $M-Z$ relation as measured by the direct method on spectra stacked by mass (\citealt{am13}; bottom left), or the SDSS ``fundamental relation'', plotting metallicity against a combination of stellar mass and SFR that minimizes scatter (\citealt{am13}; bottom right). Additional samples shown are DEEP2 star-forming galaxies \citep{hkp+05}, nearby blue compact galaxies (BCGs; \citealt{kbg+07}) and ``Green Pea'' galaxies \citep{apv+12}. The host of PS1-10bzj is consistent with nearby relations in each case, and similar to the GRB host galaxies with direct metallicity measurements.
\label{fig:mass_met} }
\end{figure*}

\section{Conclusions}
\label{sec:conc}
We show that PS1-10bzj is a hydrogen-poor superluminous supernova, spectroscopically similar to the objects described in \citet{qkk+11} and \citet{ccs+11}. Compared to previous events, it has the fastest rise time and lowest peak luminosity.  From our reconstructed bolometric lightcurve, we estimate the total energy radiated over the time period observed to be $\sim 3.5 \times 10^{50}{\rm erg}$, and the bolometric magnitude at peak to be about $-21.4$~mag. A magnetar model can fit the observed light curve, velocities and temperatures. Proposed interaction scenarios for SLSNe can also match the observed energetics but would require at least $\sim 3$ M$_{\odot}$ of hydrogen-poor circumstellar material. The lack of intermediate-width lines in the spectra, like with other SLSNe, also speaks against this model. A normal, Ni-decay Ic model would require $M_{\rm Ni} = 7$~M$_{\odot}$ and the ejecta composition to be $\gtrsim 80$ \% Ni by mass, so although PS1-10bzj shows less extreme energetics than other hydrogen-poor SLSNe, radioactive decay is unlikely to be the primary energy source.

Like SN\,2010gx and PTF09cnd, PS1-10bzj developed a number of spectral features after peak. Our model fits these features with intermediate-mass elements Mg, Ca and Si, and Fe is also likely. We do not have the spectroscopic coverage to determine whether these features at late times evolved into a more typical Type Ic SN spectrum, as was seen in the other two objects. However, PS1-10bzj is interesting in the comparison to Type Ic SNe in the sense that it extends the distribution of SLSNe towards lower luminosities. Continuing to map out the low-luminosity tail of the SLSN population will be necessary to determine whether the two classes represent truly distinct phenomena, or whether there is a smooth continuum between them. If the timescales of PS1-10bzj are typical for the lower-luminosity objects, this may present a challenge for finding such events as they will not stand out photometrically as much as higher-luminosity events and will require spectroscopic confirmation. 

The host galaxy of PS1-10bzj is detected both in our PS1 template images and in catalogs covering the ECDF-S. Combining this photometry with emission line measurements, we find that the host is a low luminosity ($M_B \simeq -18$~mag; $L \simeq 0.05L_*$), low metallicity ($Z = 0.13 Z_{\odot}$), low stellar mass ($M_* \approx 2.4 \times 10^7$~M$_{\odot}$) galaxy. It is forming stars at a rate of $\sim 2-3$~M$_{\odot}$yr$^{-1}$, resulting in a high sSFR (100 Gyr$^{-1}$). Archival {\it HST} imaging further reveal the host to be compact, with a physical diameter $\lesssim 800$ pc. While the metallicity is not as low as the host galaxy of the superluminous SN\,2010gx, the discovery of a second low metallicity host galaxy supports the hypothesis that metallicity may be important in the progenitor channel of SLSNe. Compared to the host of SN\,2010gx, the host of PS1-10bzj has a higher SFR, and is generally consistent with the $M-Z$ relation for starforming galaxies at lower redshifts \citep{mcm+10,am13}. Further increasing the sample of SLSNe with well-studied host galaxies will be necessary to assess whether this metallicity trend holds, and shed light on the nature of these extreme explosions.

\acknowledgements
We thank the staffs at PS1, Gemini and Magellan for their assistance with performing these observations. The Pan-STARRS1 Surveys (PS1) have been made possible through contributions of the Institute for Astronomy, the University of Hawaii, the Pan-STARRS Project Office, the Max-Planck Society and its participating institutes, the Max Planck Institute for Astronomy, Heidelberg and the Max Planck Institute for Extraterrestrial Physics, Garching, The Johns Hopkins University, Durham University, the University of Edinburgh, Queen's University Belfast, the Harvard-Smithsonian Center for Astrophysics, the Las Cumbres Observatory Global Telescope Network Incorporated, the National Central University of Taiwan, the Space Telescope Science Institute, and the National Aeronautics and Space Administration under grant No. NNX08AR22G issued through the Planetary Science Division of the NASA Science Mission Directorate. This work is based in part on observations obtained at the Gemini Observatory (under Programs GS-2010B-Q-4 and GS-2011A-Q-29 (PI: Berger) and GS-2011B-Q-44 (PI: Chornock)), which is operated by the Association of Universities for Research in Astronomy, Inc., under a cooperative agreement with the NSF on behalf of the Gemini partnership: the National Science Foundation (United States), the Science and Technology Facilities Council (United Kingdom), the National Research Council (Canada), CONICYT (Chile), the Australian Research Council (Australia), Ministério da Ciência, Tecnologia e Inovação (Brazil), and Ministerio de Ciencia, Tecnología e Innovación Productiva (Argentina). This paper includes data gathered with the 6.5 m Magellan Telescopes located at Las Campanas Observatory, Chile. This paper includes data based on observations made with the NASA/ESA {\it Hubble Space Telescope} and obtained from the Hubble Legacy Archive, which is a collaboration between the Space Telescope Science Institute (STScI/NASA), the Space Telescope European Coordinating Facility (ST-ECF/ESA) and the Canadian Astronomy Data Centre (CADC/NRC/CSA). Some of the computations in this paper were run on the Odyssey cluster supported by the FAS Science Division Research Computing Group at Harvard University.  The research leading to these results has received funding from the European Research Council under the European Union's Seventh Framework Programme (FP7/2007-2013)/ERC grant agreement No. [291222]  (PI: S.~J.~Smartt). Partial support for this work was provided by National Science Foundation grants AST-1009749 and AST-1211196.

{\it Facilities:} \facility{PS1}, \facility{Gemini:South}, \facility{Magellan:Baade}, \facility{Magellan:Clay}.


\begin{deluxetable}{lcccc}
\tablewidth{0pt}
\tablecaption{PS1-10bzj Photometry\label{tab:phot}}
\tablehead{ 
   \colhead{MJD} & \colhead{Phase\tablenotemark{a}} & \colhead{Filter}  &
   \colhead{AB Magnitude}  & \colhead{Telescope/Instrument}}
\startdata
      55509.4  & $ -32.9$ &  \gps  & $> 23.74$           &   PS1             \\
      55518.4  & $ -27.4$ &  \gps  & $> 22.72$           &   PS1             \\
      55524.4  & $ -23.8$ &  \gps  & $> 22.21$           &   PS1             \\
      55545.4  & $ -11.1$ &  \gps  & 21.30 $\pm$   0.12  &   PS1             \\
      55548.4  & $ -9.2 $ &  \gps  & 21.26 $\pm$   0.15  &   PS1             \\
      55557.3  & $ -3.8 $ &  \gps  & 21.33 $\pm$   0.05  &   PS1             \\
      55574.6  & $ +6.6 $ &  $g$   & 21.58 $\pm$   0.10  &   Magellan/LDSS3   \\ 
      55586.1  & $ +13.6$ &  $g$   & 22.32 $\pm$   0.06  &   Gemini-S/GMOS   \\
      55589.0  & $ +15.4$ &  $g$   & 22.83 $\pm$   0.16  &   Gemini-S/GMOS   \\
      55596.2  & $ +19.7$ &  \gps  & $> 23.43$           &   PS1             \\
      55627.5  & $ +38.7$ &  $g$   & $> 24.47$           &   Magellan/IMACS  \\ 
      55648.9  & $ +51.7$ &  $g$   &  $> 22.07 $         &   Gemini-S/GMOS   \\ \hline 
		 
      55509.4  & $ -32.9$ &  \rps  & $> 24.09 $          &   PS1             \\
      55518.5  & $ -27.4$ &  \rps  & $> 23.12 $          &   PS1             \\
      55524.4  & $ -23.8$ &  \rps  & $> 22.54 $          &   PS1             \\
      55545.4  & $ -11.1$ &  \rps  & 21.55 $\pm$ 0.06    &   PS1             \\
      55548.4  & $ -9.2 $ &  \rps  & 21.32 $\pm$ 0.10    &   PS1             \\
      55557.4  & $ -3.8 $ &  \rps  & 21.44 $\pm$ 0.06    &   PS1             \\
      55574.6  & $ +6.6 $ &  $r$   & 21.33 $\pm$ 0.05    &   Magellan/LDSS3   \\
      55586.1  & $ +13.6$ &  $r$   & 21.49 $\pm$ 0.07    &   Gemini-S/GMOS   \\
      55589.0  & $ +15.4$ &  $r$   & 21.86 $\pm$ 0.08    &   Gemini-S/GMOS   \\
      55596.2  & $ +19.7$ &  \rps  & 22.25 $\pm$ 0.08    &   PS1             \\
      55627.5  & $ +38.7$ &  $r$   & 23.33 $\pm$ 0.13    &   Magellan/IMACS  \\
      55652.9  & $ +54.1$ &  $r$   & $> 22.5 $           &   Gemini-S/GMOS   \\ \hline 

      55507.5  & $ -34.0$ & \ips   & $> 23.89$           &   PS1             \\
      55510.4  & $ -32.3$ & \ips   & $> 23.74$           &   PS1             \\		
      55516.4  & $ -28.6$ &  \ips  & $> 23.53 $          &   PS1             \\
      55534.4  & $ -17.7$ &  \ips  &  22.92 $\pm$ 0.27   &   PS1             \\
      55546.4  & $ -10.4$ &  \ips  &  21.68 $\pm$ 0.08   &   PS1             \\
      55555.4  & $ -5.0 $ &  \ips  &  21.51 $\pm$ 0.07   &   PS1             \\
      55574.6  & $ +6.6 $ &  $i$   &  21.37 $\pm$ 0.07   &   Magellan/LDSS3   \\
      55576.3  & $ +7.7 $ &  \ips  &  21.49 $\pm$ 0.09   &   PS1             \\
      55586.1  & $ +13.6$ &  $i$   &  21.48 $\pm$ 0.03   &   Gemini-S/GMOS   \\
      55588.2  & $ +14.9$ &  \ips  &  21.78 $\pm$ 0.08   &   PS1             \\
      55594.2  & $ +18.5$ &  \ips  &  21.80 $\pm$ 0.07   &   PS1             \\
      55597.2  & $ +20.3$ &  \ips  &  21.78 $\pm$ 0.07   &   PS1             \\
      55627.5  & $ +38.7$ &  $i$   &  22.68 $\pm$ 0.13   &   Magellan/IMACS  \\
      55648.9  & $ +51.7$ &  $i$   &  $> 22.19 $         &   Gemini-S/GMOS   \\ \hline 
		 
      55508.4  & $ -33.5$ & \zps   &  $> 23.56$          &   PS1             \\
      55511.4  & $ -31.7$ & \zps   &  $> 23.13$          &   PS1             \\
      55517.4  & $ -28.0$ & \zps   &  $> 23.28$          &   PS1             \\
      55547.3  & $ -9.9 $ & \zps   &  21.71 $\pm$ 0.07   &   PS1             \\
      55568.3  & $ +2.8 $ & \zps   &  21.42 $\pm$ 0.08   &   PS1             \\
      55574.6  & $ +6.6 $ & $z$    &  21.28 $\pm$ 0.14   &   Magellan/LDSS3   \\
      55577.3  & $ +8.3 $ &  \zps  &   21.57 $\pm$ 0.11   &   PS1             \\
      55586.1  & $ +13.6$ &  $z$   &  21.59 $\pm$ 0.22   &   Gemini-S/GMOS   \\
      55586.3  & $ +13.7$ &  \zps  &  21.55 $\pm$ 0.08   &   PS1             \\
      55589.2  & $ +15.5$ &  \zps  &  21.57 $\pm$ 0.08   &   PS1             \\
      55595.2  & $ +19.1$ &  \zps  &  21.84 $\pm$ 0.09   &   PS1             \\
      55627.5  & $ +38.7$ &  $z$   &  22.32 $\pm$ 0.12   &   Magellan/IMACS  
\enddata
\tablenotetext{a}{In rest-frame days, relative to maximum light on MJD 55563.65}      
\end{deluxetable}

\begin{deluxetable*}{lcccccccc}
\tabletypesize{\scriptsize}
\tablecaption{Log of Spectroscopic Observations}
\tablehead{
\colhead{UT Date} &
\colhead{Epoch\tablenotemark{a}} &
\colhead{Instrument} &
\colhead{Wavelength Range} &
\colhead{Slit} &
\colhead{Grating} &
\colhead{Filter} &
\colhead{Exp. time} &
\colhead{Mean} \\
\colhead{(YYYY-MM-DD.D)} &
\colhead{(days)} &
\colhead{} &
\colhead{(\AA)} &
\colhead{($\arcsec$)} &
\colhead{} &
\colhead{} &
\colhead{(s)} &
\colhead{Airmass}
}
\startdata
2011-01-13.2 &  6.7 & LDSS3  & 3540$-$9450 & 0.75 & VPH-all & none & 3900 & 1.3 \\
2011-01-25.1 & 13.9 & GMOS-S & 3320$-$6140 & 1.0 & B600 & none & 2400 & 1.4 \\
2011-01-28.1 & 15.7 & GMOS-S & 5890$-$10100 & 1.0 & R400 & OG515 & 3000& 1.1 \\
2011-04-02.0 & 54.5 & GMOS-S & 5530$-$9830 & 1.0 & R400 & OG515 & 900 & 1.9 \\
2011-04-03.0 & 55.1 & GMOS-S & 5530$-$9830 & 1.0 & R400 & OG515 & 1800& 1.9 
\enddata
\tablenotetext{a}{In rest-frame days relative to maximum light on UT 2011-01-02.7.}
\label{tab:spec}
\end{deluxetable*}

\begin{deluxetable}{cccc}
\tablewidth{0pt}
\tablecaption{PS1-10bzj Host Galaxy Photometry\label{tab:host}}
\tablehead{
 \colhead{UT Date}    &\colhead{Filter} &
\colhead{AB Magnitude}   & \colhead{Telescope/Instrument}}
\startdata
        & \gps     & 24.37 $\pm$ 0.13 &  PS1              \\
        & \rps     & 24.00 $\pm$ 0.12 &  PS1              \\
        & \ips     & 23.76 $\pm$ 0.10 &  PS1              \\
        & \zps     & 22.73 $\pm$ 0.05 &  PS1              \\
        & \yps     &     $> 21.7 $    &  PS1              \\
 2011-11-29      & $g'$      & 24.37 $\pm$ 0.08 &  Gemini-S/GMOS    \\
 2011-10-21      & $r'$      & 23.86 $\pm$ 0.18 &  Magellan/LDSS3   \\
 2011-09-20      & $i'$      & 23.12 $\pm$ 0.07 &  Gemini-S/GMOS    \\
 2012-07-19      & $z'$     & 23.67 $\pm$ 0.15 &  Magellan/IMACS   \\
 2012-12-04      & $J$      &     $> 23.8$     &  Magellan/FourStar\\
 2011-12-07      & $K$      &     $> 22.7$     &  Magellan/FourStar\\
     & F606W    & 24.13 $\pm$ 0.05 &  HST/ACS\tablenotemark{a} \\
     & F850LP   & 23.63 $\pm$ 0.06 &  HST/ACS\tablenotemark{a} \\
                 &  $U38$   & 24.89 $\pm$ 0.08 &  ESO MPG 2.2m/WFI\tablenotemark{b} \\
                 &  $U$     & 24.86 $\pm$ 0.04	&  ESO MPG 2.2m/WFI\tablenotemark{b} \\
                 &  $B$     & 24.45 $\pm$ 0.02	&  ESO MPG 2.2m/WFI\tablenotemark{b} \\
                 &  $V$     & 24.44 $\pm$ 0.02	&  ESO MPG 2.2m/WFI\tablenotemark{b} \\
                 &  $R$     & 24.22 $\pm$ 0.02	&  ESO MPG 2.2m/WFI\tablenotemark{b} \\
                 &  $I$     & 23.23 $\pm$ 0.05	&  ESO MPG 2.2m/WFI\tablenotemark{b} \\
                 &  $z'$     & 23.39 $\pm$ 0.13 &  CTIO 4m/Mosaic-II\tablenotemark{b} \\
                 & IA427    & 24.25 $\pm$ 0.07 &  Subaru/Suprime-Cam\tablenotemark{c}  \\
 		 & IA445    & 24.49 $\pm$ 0.07	&  Subaru/Suprime-Cam\tablenotemark{c}  \\
		 & IA464    & 24.59 $\pm$ 0.14	&  Subaru/Suprime-Cam\tablenotemark{c}  \\
		 & IA484    & 24.45 $\pm$ 0.03	&  Subaru/Suprime-Cam\tablenotemark{c}  \\
		 & IA505    & 24.42 $\pm$ 0.06	&  Subaru/Suprime-Cam\tablenotemark{c}  \\
		 & IA527    & 24.53 $\pm$ 0.03	&  Subaru/Suprime-Cam\tablenotemark{c}  \\
		 & IA550    & 24.41 $\pm$ 0.05	&  Subaru/Suprime-Cam\tablenotemark{c}  \\
		 & IA574    & 24.42 $\pm$ 0.06	&  Subaru/Suprime-Cam\tablenotemark{c}  \\
		 & IA598    & 24.21 $\pm$ 0.02	&  Subaru/Suprime-Cam\tablenotemark{c}  \\
		 & IA624    & 23.84 $\pm$ 0.02	&  Subaru/Suprime-Cam\tablenotemark{c}  \\
		 & IA651    & 24.08 $\pm$ 0.02	&  Subaru/Suprime-Cam\tablenotemark{c}  \\
		 & IA679    & 24.59 $\pm$ 0.03	&  Subaru/Suprime-Cam\tablenotemark{c}  \\
		 & IA709    & 24.64 $\pm$ 0.12	&  Subaru/Suprime-Cam\tablenotemark{c}  \\
		 & IA738    & 24.58 $\pm$ 0.04	&  Subaru/Suprime-Cam\tablenotemark{c}  \\
		 & IA767    & 24.62 $\pm$ 0.10	&  Subaru/Suprime-Cam\tablenotemark{c}  \\
		 & IA797    & 23.78 $\pm$ 0.06	&  Subaru/Suprime-Cam\tablenotemark{c}  \\
		 & IA827    & 22.25 $\pm$ 0.03	&  Subaru/Suprime-Cam\tablenotemark{c}  \\
		 & IA856    & 24.68 $\pm$ 0.16 &  Subaru/Suprime-Cam\tablenotemark{c}  
\enddata
\tablenotetext{a}{Data from GEMS survey catalog \citep{rbb+04}}
\tablenotetext{b}{Data from GaBoDs survey catalog \citep{tfd+09}}
\tablenotetext{c}{Data from MUSYC survey catalog \citep{cdu+10}}
\end{deluxetable}

\begin{deluxetable}{lccc}
\tablewidth{0pt}
\tablecaption{Host Galaxy Emission Line Fluxes\label{tab:emlines}}
\tablehead{
\colhead{Line}   & \colhead{} & \colhead{Flux ($10^{-16}$ erg s$^{-1}$ cm$^{-2}$)} 
 & \colhead{}\\
 \colhead{}&  \colhead{Jan 13} &
\colhead{Jan 28} & \colhead{Apr 3} }
\startdata
  $[$\ion{O}{3}$]\lambda5007$   & 6.43 $\pm$ 0.18   &  6.36 $\pm$ 0.07  & 6.20 $\pm$ 0.08 \\
   $[$\ion{O}{3}$]\lambda4959$  & 1.89 $\pm$ 0.14 &  2.04 $\pm$ 0.05  & 2.32 $\pm$ 0.06 \\
   H$\beta$                     & 1.05 $\pm$ 0.16 &  1.04 $\pm$ 0.05  & 0.94 $\pm$ 0.08 \\
   $[$\ion{O}{3}$]\lambda4363$  &  \nodata         &  0.14 $\pm$ 0.04  & \nodata      \\
   H$\gamma$	                & 0.47 $\pm$ 0.07 &  0.53 $\pm$ 0.04  & 0.43 $\pm$ 0.05 \\
   H$\delta$                    & 0.20 $\pm$ 0.06 &  0.22 $\pm$ 0.04  & 0.24 $\pm$ 0.04 \\
   H$\epsilon$ + $[$\ion{Ne}{3}$]\lambda3968$		& 0.28 $\pm$ 0.08 &  0.27 $\pm$ 0.05  &	0.28 $\pm$ 0.06	\\
   H$\zeta$			& \nodata	  &  0.19 $\pm$ 0.05  &	\nodata		\\
   $[$\ion{Ne}{3}$]\lambda3869$	& 0.48 $\pm$ 0.08 &  0.37 $\pm$ 0.04  & 0.58 $\pm$ 0.06	\\
   $[$\ion{O}{2}$]\lambda3727$  &  0.97 $\pm$ 0.09 &  0.83 $\pm$ 0.06  & 1.07 $\pm$ 0.08 
\enddata
\end{deluxetable}

\end{document}